\documentclass[12pt]{article}
\usepackage[utf8]{inputenc}

\usepackage{amsmath}
\usepackage{amsfonts}
\usepackage{amssymb}
\usepackage{makeidx}
\usepackage{graphicx}
\usepackage{subcaption}
\usepackage{listings}
\usepackage{apacite}
\usepackage{natbib}
\usepackage{indentfirst}
\usepackage{caption}
\usepackage{enumitem}
\usepackage{ulem}

\usepackage{ulem}
\usepackage{soul}
\usepackage{color}

\usepackage{mathtools}
\usepackage{tabularx}
\usepackage{siunitx}
\usepackage{multirow}
\usepackage{booktabs}
\newcolumntype{Y}{>{\centering\arraybackslash}X}

\usepackage{setspace}
\usepackage{authblk}
\usepackage[pdfborder={0 0 0}]{hyperref}

\title{\bf A New Set of Financial Instruments}
\author[a]{Abootaleb Shirvani}
\author[b]{Stoyan V. Stoyanov}
\author[c]{Svetlozar T. Rachev}
\author[d]{Frank J. Fabozzi}

\begin{spacing}{.6}
	\affil[a]{\small Department of Mathematics and Statistics, Texas Tech University
		\url{abootaleb.shirvani@ttu.edu}}
	\affil[b]{\small College of Business, Stony Brook University\\
		\url{stoyan.stoyanov@stonybrook.edu}}
	\affil[d]{\small Department of Mathematics and Statistics, Texas Tech University\\
		\url{zari.rachev@ttu.edu}}
	\affil[c]{\small EDHEC Business School\\
		\url{frank.fabozzi@edhec.edu}}	
\end{spacing}

\begin{document}
\thispagestyle{plain}

\date{}

\maketitle
\begin{spacing}{0.90}
\noindent \textbf{Abstract}\ \ \ \ \ 
In complete markets there are risky assets and a riskless asset.  It is assumed that the riskless asset and the risky asset are traded continuously in time and that the market is frictionless.  In this paper, we propose a new method for hedging derivatives assuming that a hedger should not always rely on trading existing assets that are used to form a linear portfolio comprised of the risky asset, the riskless asset, and standard derivatives, but rather should design a set of specific, most-suited financial instruments for the hedging problem. We introduce a sequence of new financial instruments best suited for hedging jump-diffusion and stochastic volatility market models. The new instruments we introduce are perpetual derivatives. More specifically, they are options with perpetual maturities. In a financial market where perpetual derivatives are introduced, there is a new set of partial and partial-integro differential equations for pricing derivatives. Our analysis demonstrates that the set of new financial instruments together with a risk measure called the tail-loss ratio measure defined by the new instrument's return series can be potentially used as an early warning system for a market crash.
\end{spacing}

\begin{spacing}{1.00}
\newpage
\section{Introduction}
\noindent In complete markets there are risky assets (such as stocks) and a riskless asset (such as government-guaranteed bond).  The risky asset and the riskless asset are traded continuously in time and the market is assumed to be frictionless. These assets are referred to as “basic assets”. Given a derivative with an underlying assets being the risky and riskless asset, the derivative price is spanned over the risky asset’s prices and the riskless asset’s price. That is, contingent pricing theory tells us that the derivative price process is replicated by the price process of a self-financing portfolio consisting of these two existing assets. 

In this paper we introduce new risky assets which we propose as trading assets (i.e., available for trade continuously in time).  The proposed new risky assets will be new assets if they are accepted by the market as tradable assets. These new assets are perpetual options (i.e., options with perpetual maturities). Because the proposed perpetual derivatives incorporate information about the underlying risky asset, a riskless asset, and the underlying risky asset’s volatility, they could potentially be of greater interest to market participants than the underlying asset itself. In a way, our perpetual financial instruments are portfolios driven by the underlying risky asset, and we suggest it as a potentially tradable product. We believe that introducing this new set of financial instruments will contribute to enhancing market efficiency. Since real markets are not complete, the introduction of the proposed financial instruments would benefit real financial markets.

Of course, a natural question is that in complete markets there are already contingent claims traded (calls and puts), so why is it necessary to have new financial instruments for spanning? The answer is simple. This is because existing traded options have a time to maturity and their value depends critically on the time to maturity. A preferred hedging instrument should only depend on (1) the current time, (2) current value of the underlying asset, (3) the prevailing riskless interest rate, and (4) the current value of the asset’s volatility. It should not depend on the option's time to maturity. That is the critical feature of the set of new financial instrument we propose here, options with infinite maturity. We believe that the new financial instruments we propose in this paper will motivate market participants to recommend to exchanges to design such contracts to help facilitate more efficient trading.  

In creating the new financial instruments, we use the basic framework that any asset in a complete market can be replicated by existing assets. In the classical Black-Scholes-Merton (BSM) model \citep[see][]{Black:1973, Merton1973b}, the price of a risky asset is driven by geometric Brownian motion and the riskless asset. In the classical BSM model, one set of basic assets is the riskless asset and the underlying risky asset. In this paper, we will show that our new financial instrument is a basic asset. The new financial instrument, together with the riskless asset and the underlying risky asset, span any derivative in the BSM model. 

So our new financial instrument can and should be replicated by basic assets. Thus, if the new financial instrument is available for trade as is the underlying asset, the seller of our perpetual derivative can form a replicating portfolio so that a trader is constantly able to hedge the risk exposure by investing in the risky asset. That is why we view our new financial instrument (as any risky asset in a complete market) both as an asset to be bought (taking a long position) and at the same time to be sold (and thus replicated using  a perfect hedged replicating portfolio).

Extensions of the classical BSM framework allow for stochastic volatility and jump risk in the price process which make he market incomplete. For example, \cite{Merton:1976} introduces a jump-diffusion model and uses a short cut by averaging the jump risk noting that his model
is not equivalent to an equilibrium model.
Yet, we show that Merton’s model becomes complete if we introduce a special bond with a maturity being the first arrival (default time) of a Poisson process jump. In our paper we continue Merton’s line of research by discussing how this new special bond should be used for hedging and in doing so we derive a new partial differential equation for any contingent claim that will make the market complete. 

With respect to volatility markets, such markets are incomplete markets in the sense that they too introduce two sets of risk: price risk and volatility risk. However, if the volatility of the asset is traded, the local volatility market becomes complete. Furthermore, if the volatility of volatility is traded, then we have four traded assets (underlying risks asset, volatility, volatility of volatility, and riskless asset) forming a complete market within the local volatility market model. Currently market participants employ the CBOE's VIX \footnote{ VIX is an index created by CBOE, representing 30-day implied volatility calculated by S\&P500 options (see \url{http://www.cboe.com/vix}).} (as volatility of SPDR \footnote{See SPDR S\&P 500 ETF Indices, \url{https://us.sprdrs.com/}.}) and VVIX \footnote{The VVIX is an index created by the CBOE, (see http://www.cboe.com/products/vix-index-volatility/volatility-on-stock-indexes/the-cboe-vvix-index/vvix-whitepaper). It is a volatility of volatility (vol-of-vol) measure, and represents 30-day implied volatility calculated from  VIX options.} (as volatility of VIX) as traded assets (assets available for trade). In our paper we derive the partial differential equations for derivatives that will make the market complete.

Our general methodology, namely, to search for hedging instruments that are most suited for the hedging problems is applied to three classical problems associated with option pricing models. First, we apply our model to continuous-time BSM markets. We derive a new set of perpetual derivatives which can be as a potentially tradable product. We also extend this approach to multi-asset markets (following general multivariate It\^{o} processes). Further, we explain and evaluate market risk before and during the recent distressed market period by using the new perpetual derivative.  We demonstrate that the new perpetual derivative, together with a risk measure which we refer to as the tail-loss ratio (TLR), was capable of explaining and evaluating market risk before and during the potential distressed market period.  We present the TLR index for the new perpetual derivative and the SPDR S\&P 500  index for period from 2000-2018.  We assess the forecasting performance of the new perpetual derivative by comparing the TLR index with the TLR index for the SPDR S\&P 500 index. Our empirical evidence  suggests that the TLR derived by the new financial instrument we propose performs well in predicting a real-world market crash.


The second problem associated with option pricing models that we tackle in this paper deals with hedging in Merton's jump-diffusion option pricing model \citep[see][]{Merton:1976}. Here the classical approach is to use a riskless asset and the stock, but since these two instruments cannot be used to hedge jump risk, this risk is left unhedged. Now we again apply our general framework to answer the following question: What kind of tradable (possibly non-linear) financial instrument is best suited for designing a hedging strategy to eliminate jump risk?  Following the approach by \cite{Runggaldier:2003}, we answer the question and thus derive an analogue of Merton's partial integro-differential equation (PIDE) for the derivative price of a fully hedged portfolio.

Our third application of the proposed approach is hedging in the presence of stochastic volatility. Here there are two sources of uncertainty:  market risk and volatility risk. Although market risk is readily hedged by trading the underlying asset, attempting to hedge volatility risk requires an additional derivative with a longer maturity than the option that the hedger seeks to hedge. This approach creates a vicious circle in that the hedger is trying to hedge an option using another option for which the hedger does not know its contract value. Thus, an analogue of the market risk premium is introduced (generally understood as the volatility risk premium), which now enters the model as a parametric function which should be potentially calibrated. Instead, following the approach suggested by \cite{Davis:2004}, we take a different approach by posing the question of the most suitable tradable instrument for that hedging problem and describe the nature of the volatility risk premium by applying the Consumption Capital Asset Pricing Model (CCAPM) formulated by \cite{Breeden:1979}. 

This is the essence of our methodology: in an incomplete market, we do not leave the risk premia from different risk factors as unknown functions (which should be eventually estimated or calibrated). Rather we select the best suited financial instruments, that should be introduced as publicly traded assets. These new assets should make the underlying market complete. Furthermore, even if the market is complete, we identify various spanning bases of assets, best suitable for the hedging problem under consideration. 

We can sum up our methodology based on the following idea: Every hedging problem has its own set of most-suited (``ideal'') set of hedging instruments. There is no universal hedging instrument, as there is no universal hedging problem.\footnote{It is interesting to note that the renowned Russian mathematician Andrey N. Kolmogorov used to comment that every approximation problem in functional analyses and probability theory requires a specially designed distance measure (best-suited metric) in its solution \citep[][]{Rachev:2013}. Similarly, in hedging problems, the choice of the hedging instruments, should not necessarily be the standard ones (the stock and the riskless asset), but the ones that best reflect the nature of the hedging problem.} Furthermore, ideal hedging instruments cannot make a bad model good; an intrinsically bad model cannot be made acceptable with any enhancements. 

The paper is organized as follows. In the next section, we introduce a new set of perpetual derivatives that will serve as the basis assets in hedging portfolios. In Section 3, we apply our general methodology to determine what kind of tradable financial instrument is best suited for the problem of eliminating the jump risk in Merton's jump-diffusion model. The solution is to use a new financial instrument, which can be viewed as an interest-bearing bond where the interest payments occurr at Poisson arrivals. $\ $The next application of the general method is hedging within a stochastic volatility model. In Section 4 we show that volatility indexes should be used as desirable hedging instruments. In all applications, the corresponding analogues of the Black-Scholes and Merton's equations are derived. In Section 5 the TLR indices of the new perpetual derivative and SPDR S\$P 500  (an exchange-traded fund) index are presented and the results then compared. The proofs are given in the Appendix.

\section{A class of ``ideal" perpetual derivatives}

\noindent Consider the classical Black-Scholes-Merton BSM-framework

\noindent (a)  a risky asset (stock) with price dynamics given by
\begin{equation}
	\label{a2_1}
	{dS}_t=\mu S_tdt+\ \sigma S_t{dB}_t,t\ge 0,\  {\ }S_0>0,\  \mu {>0,\ }\sigma >0 {\ };     
\end{equation}
on a stochastic basis $\left( {\Omega },\mathcal{F},{\left\{{\mathcal{F}}_t\right\}}_{t\ge 0},\mathbb{P}\right)$  \footnote{ $\left( {\Omega },\mathcal{F},{\left\{{\mathcal{F}}_t\right\}}_{t\ge 0} {,}\mathbb{P}\right)$ is generated by the Brownian motion  $B_t,t\ge 0$}  representing the natural world, $ \mu $ is the instantaneous stock's mean return, and $\ \sigma $ is the stock's volatility;\\
(b) a riskless bond given by  

\begin{equation} 
	\label{a2_2} 
	d\beta_{t}= {r}{\beta }_tdt,t\ge 0, \,\beta_0=1,  r>0,
\end{equation}             
where $r$ is the risk-free rate;\\ 
(c) A European contingent claim (ECC) with price process ${ {Y}}_{ {t}}=Y\left(S_t,t\right)$ at $t\in \left[0,T\right], $ maturity $T$,  terminal value ${ {Y}}_{ {T}}={\mathcal{G}(S}_T)$, and price dynamics given by the It\^{o} process:

\begin{equation} 
	\label{a2_3}
	d{ {Y}}_{ {t}}=\left(\frac{\partial Y\left(S_t,t\right)}{\partial t}+\ \frac{\partial Y\left(S_t,t\right)}{\partial x}\  \mu S_t+\frac{1}{2}\frac{{\partial }^2Y\left(S_t,t\right)}{\partial x^2}\ {\sigma }^2S^2_t\right)dt+\frac{\partial Y\left(S_t,t\right)}{\partial x}\sigma S_t{dB}_t. 
\end{equation} 

Under the equivalent martingale measure (EMM)  $\mathbb{Q} {\sim }\mathbb{P}$ the discounted price process $\frac{{ {Y}}_{ {t}}}{{\beta }_t}$  is a martingale. On the real world $\mathbb{P},\ $the derivative is hedged by a self-financing strategy$\ \ Y\left(S_t,t\right)=a_tS_t+b_t{\beta }_t$, with $a_t=\ \frac{\partial Y\left(S_t,t\right)}{\partial x}$. The riskless bond and the risky asset are not the only tradable assets that can be used to replicate the ECC price-dynamics. In hedging the ECC, the trader could use a perpetual derivative available for trading as shown in the next proposition:

\noindent \textit{Proposition 2.1.} Let $\varsigma \in R$ be a parameter and ${\mathbb{V}}^{\left(\varsigma \right)}$ is designated as the risky asset with price process  $V^{\left(\varsigma \right)}_t={S_t}^{\varsigma }{\beta }^{\gamma }_t,\ t\ge 0$, where $\gamma =\frac{1-\zeta }{r}\left(r+\frac{1}{2}\zeta {\sigma }^2\right)$. Then the price process  $V^{\left(\varsigma \right)}_t,t\ge 0$, discounted by the riskless bond rate is a martingale under the EMM $\mathbb{Q} {\sim }\mathbb{P}$, and thus, security ${\mathbb{V}}^{\left(\varsigma \right)}$ can be traded within the BSM market model \eqref{a2_1} and \eqref{a2_2}  (see Appendix A.1 for the proof).

Of interest is the perpetual derivative   ${\mathbb{V}=\mathbb{V}}^{\left(\delta \right)}$, where $\delta =\frac{-2r}{{\sigma}^2}$, with price process   $V_t=V^{\left(\delta\right)}_t={S_t}^{\delta}$, which arises with $\gamma = 0$ and is not a function of the bond price.  Currently the existing ``basic" traded assets are a bond (designated as basic asset of order 0, shortly  ${\mathbb{W}}^{\left(0\right)}$) and the risky asset (designated as basic asset of order 1, shortly, ${\mathbb{W}}^{\left(1\right)}$). For a given $\varsigma \in R$ , let  ${\mathbb{W}}^{\left(\varsigma\right)}$ be a derivative with price process  $W^{\left(\varsigma\right)}_t:=V^{\left(\varsigma\right)}_t$ defined in Proposition 2.1. Then ${\mathbb{W}}^{\left(\varsigma\right)},\ \varsigma\in {\mathbb{R}}^{\left(*\right)}$ \footnote{We denote $\mathbb{R}\coloneqq \left(-\infty,\infty \right) $ and ${\mathbb{R}}^{\left(*\right)} {=}\mathbb{R}\cup \left\{\infty \right\}$.} can be publicly traded, as $W^{\left(\varsigma\right)}_t$ discounted by the risk-free rate will be a $\mathbb{Q}$-martingale.  We designate ${\mathbb{W}}^{\left(\varsigma\right)}$ as a basic asset of order $\varsigma$. 


Consider now the multidimensional case \footnote{See Sections 5.I and 6.I in \cite{Duffie:2001} for the regularity and market completeness conditions.}: ${\mathbb{B}}_t={\left(B^{\left(1\right)}_t,\dots ,B^{\left(d\right)}_t\right)}^T,\ t\ge 0$ is a $d$-dimensional standard Brownian motion and the $d$-dimensional price process ${\ \mathbb{S}}_t={\left(S^{\left(1\right)}_t,\dots ,S^{\left(d\right)}_t\right)}^T$
is an It\^{o}-process with 

\begin{equation} 
	\label{a2_4}
	\left\lbrace   \begin{array}{ll}
		{dS}^{\left(i\right)}_t={\mu }^{\left(i\right)}_tS^{\left(i\right)}_tdt+\sum^d_{j=1}{{\sigma }^{\left(i,j\right)}_t}S^{\left(i\right)}_t{dB}^{\left(j\right)}_t,\ i=,1\dots d \\ 
		{\mu }^{\left(i\right)}_t={\mu }^{\left(i\right)}_{\ }\left({\ \mathbb{S}}_t {,t}\right),\,\,\,\,\,\,\,\, {\sigma }^{\left(i,j\right)}_t={\sigma }^{\left(i,j\right)}_{\ }\left({\ \mathbb{S}}_t {,t}\right), \end{array}
	\right.
\end{equation}
on a stochastic basis $\left( {\Omega },\mathcal{F},{\left\{{\mathcal{F}}_t\right\}}_{t\ge 0} {,}\mathbb{P}\right)$ generated by ${\mathbb{B}}_t,\ t\ge 0$. The market ${(\mathbb{S}}_t,$ ${\beta }_t),$ with bond
price 
\begin{equation} 
	\label{a2_5}
	{d\beta }_t={ {r}}_t{\beta }_tdt,\,t\ge 0, {\ }{\beta }_0>0, \,r_t= {r}\left({\mathbb{S}}_t {,t}\right),
\end{equation}
is assumed complete. 

\textit{Proposition 2.2} Assume that \eqref{a2_4} and \eqref{a2_5} hold. Let $V^{\left(i\right)}_t={\left(S^{\left(i\right)}_t\right)}^{{\delta }^{(i)}_t}$ $t\ge 0$, $i=1,\dots d$, then $V^{\left(i\right)}_t$ $t\ge 0$ is a risky asset, that could be publicly traded, if and only if ${\delta }^{\left(i\right)}_t=-\frac{2r_t}{\sum^d_{j=1}{{\left({\sigma}^{\left(i,j\right)}_t\right)}^2}}$. The self-financing replication of $V^{\left(i\right)}_t$ is given by ${V}^{\left(i\right)}_t=a^{\left(i\right)}_tS^{\left(i\right)}_t+{b^{\left(i\right)}_i\beta }_t $, where $a^{\left(i\right)}_t={\delta }^{(i)}_t{\left(S^{\left(i\right)}_t\right)}^{{\delta }^{\left(i\right)}_t-1}$  (see Appendix A.2 for the proof).

\section{ Eliminating jump risk in Merton's jump-diffusion pricing model}

Consider Merton's jump diffusion model \citep{Merton1973a} in which there is a stock and riskless bank account with stock price dynamics  
\begin{equation} 
	\label{a3_1}
	{dS}_t= {(}\alpha -\lambda \kappa )S_{t-}dt+\ \sigma S_{t-}{dB}_t+\left({ {y}}_t-1\right)S_{t-}dN_t,\ t\ge 0,                            
\end{equation}
defined on a stochastic basis $\left( {\Omega },\mathcal{F},{\left\{{\mathcal{F}}_t\right\}}_{t\ge 0},\mathbb{P}\right)$ generated by 

\begin{itemize}
	\item  a standard Brownian motion $B_t$ $t\ge 0$; 
	\item a homogeneous Poisson process $N_t$ $t\ge 0$, with intensity $\lambda $, and 
	\item  independent of $B_t$, $t\ge0$, and $N_t$ $t\ge 0$, independent identically distributed jumps of size 
	\noindent $J^{\left(l\right)}\triangleq J\triangleq \left\{ \begin{array}{c} ln\psi \ \ \ \ w.p.\ \ p\in (0,1) \\ 
	0\ \ \ \ \ \ \ w.p.\ \ \ \ \ 1-p \end{array}
	\right.\ ,\ l\in \mathbb{N}$ \footnote{$``\triangleq ''$, stands for ``equal in distribution", $\mathbb{N}\coloneqq \left\{1,2,\dots \right\}$.}, that is, in \eqref{a3_1},
\end{itemize}

\begin{equation} 
	\label{a3_2}
	{ {y}}_t-1\triangleq  {y}-1\triangleq \left\{\begin{array}{c}
		\psi -1\ \ \  w.p.\ \ \ \ \ \ \  p \\ 
		0\ \ \ \ \ \ \ w.p.\ \ 1-p \end{array}
	\right. 
\end{equation}
where ${ {y}}_t={ {y}}^{\left(l\right)},\ t\in \left[{\tau }^{\left(l\right)},\ {\tau }^{\left(l+1\right)}\right),$ ${\tau }^{\left(l\right)}\coloneqq { {inf} \left\{t:N_t=l\right\}},l\in \mathbb{N}$, and  ${\mathbb{E}}^{\mathbb{P}}\left({ {y}}^{\left(l\right)}-1\right)=\left(\psi -1\right)p=:\kappa$.
The riskless bank account dynamics are given by equation \eqref{a2_2}.

Consider a ECC with a contract value $Y_t=Y\left(S_t,t\right)$ at $t\in \left[0,T\right],$ maturity $T$, and terminal value $Y_T={\mathcal{G}(S}_T)$.

\noindent \textit{Proposition 3.1}. \citep{Merton:1976}: $Y\left(x,t\right)$, $x>0$, $t\in \left[0,T\right)$ satisfies the following jump-diffusion PIDE  (see Appendix A.3 for the proof)
\begin{equation}
	\label{a3_3} 
	\begin{array}{cc}
		\frac{\partial Y\left(x,t\right)}{\partial t}+\frac{1}{2}x^2{\sigma }^2\frac{{\partial }^2Y\left(x,t\right)}{\partial x^2}+rx\frac{\partial Y\left(x,t\right)}{\partial x}-rY\left(x,t\right)+\\ 
		\lambda{\mathbb{E}}^{\mathbb{P}}\left[Y\left( {y}x,t\right)-Y\left(x,t\right)\right]
		-\lambda {x\frac{\partial Y\left(x,t\right)}{\partial x}}E^{\mathbb{P}}\left[y-1\right]=0.   
	\end{array}  
\end{equation}

We derive \eqref{a3_3} by applying the CCAPM as a simple illustration of the methodology that we will apply in more general cases as shown in the Appendix. We believe that this proof is well-known, but could not find it in the literature.

In Proposition 3.1 the jump-risk is left unhedged. This is because Merton's jump-diffusion framework lacks a security that is publicly available for trading (designated as  ${\mathcal{M}}^{\left(m\right)},\ m>0)$ with which the market will be pricing the jump-occurrences. 

Here we again illustrate our approach to achieving market completeness:
$\left(i\right)$ find the most suitable hedge-instruments in the market for the hedging problem under consideration, and 
$\left(ii\right)\ $if the market does not provide such special hedging instruments, then introduce them as publicly traded assets and let the market price them.

Following the approach in \cite{Runggaldier:2003}, we introduce the dynamic price process of  ${\mathcal{M}}^{\left(m\right)}$ as a pure jump process with drift $m$:
\begin{equation}
	\label{a3_4}
	{dM}_t=mM_{t-}dt+\ \left({ {y}}_t-1\right)M_{t-}dN_t,\ t\ge 0.
\end{equation}

We view ${\mathcal{M}}^{(r)}$ as a ``riskless bond with jumps". To remove portfolio's jump-risk, the trader should form a jump-free self-financing portfolio ${\mathcal{P}}^{\left(0\right)}$ with price process  ${\mathcal{P}}_t=M_{t-}S_t-\ S_{t-}M_t$  and price dynamics given by   $d{\mathcal{P}}_t=M_{t-}{dS}_t-\ S_{t-}dM_t=(\alpha -\lambda \kappa -m)S_tM_{t-}dt+\ \sigma S_t{M_{t-}dB}_t$.

Having the bond, the stock and the new asset ${\mathcal{P}}^{\left(0\right)}$, a trader could apply Merton's Intertemporal Capital Asset Pricing Model (ICAPM) in search of optimizing his or her wealth portfolio \citep[see][section 9B]{Duffie:2001}.

Consider then an ECC written against the stock, with price process $Y_t=Y\left(S_t,M_t,t\right)$, $0\le t\le T$, terminal value ${ {Y}}_{ {T}}=Y\left(S_T,y,\ T\right)={\mathcal{G}(S}_T)$, for all $y\ge 0$. We assume that $Y\left(x,y,t\right)$, $x\ge 0$, $y\ge 0$ is (i) sufficiently smooth with respect to $x$, and $t$, and (ii) with respect to $y$, $Y\left(x,y,t\right)$, $y>0$, is left-continuous with right limits as well as continuously differentiable in the points of $y$-continuity.

\noindent \textit{Proposition 3.2.} Under assumptions \eqref{a3_1},\eqref{a3_3}, and\eqref{a3_4}, $Y\left(x,y,t\right)$, $x>0$, $y>0$, satisfies the following differential equation with solutions having jumps  (see Appendix A.4 for the proof):
\begin{equation}
	\label{a3_5}
	\begin{array}{ll}
		\frac{\partial Y\left(x,y,t\right)}{\partial t}-r\frac{\partial Y\left(x,y,t\right)}{\partial x}x- rY\left(x,y,t\right)+\frac{1}{2}\frac{{\partial }^2Y\left(x,y,t\right)}{\partial x^2} + 
		\left(r-m\right)\left\{x+y_{(-)}\frac{\partial Y\left(x,y-,t\right)}{\partial y}\right\rbrace \\- \left(r-m\right)\left\lbrace \frac{\partial Y\left(S_{t-},M_t,t\right)}{\partial x}\left[Y\left(S_t,y, t\right)-Y\left(S_{t-},y_{\left(-\right)},t\right)\right]-rY\left(S_t,y_t,t\right)\right\}=0.
	\end{array}
\end{equation}
For $r=m$, \eqref{a3_5} becomes the BSM-equation for all $y>0$. 

As a corollary of Proposition 3.2, it follows that asset  $\mathcal{M} {=}{\mathcal{M}}^{\left( {r}\right)}$ with price process: ${ {d}\mathcal{M}}_{ {t}} {=r}{\mathcal{M}}_{ {t-}} {dt+}\left({ {y}}_{ {t}} {-} {1}\right){\mathcal{M}}_{ {t-}} {d}{ {N}}_{ {t}}$ $t\ge0$  could be introduced as a publicly traded asset, along with  ${\mathbb{V}=\mathbb{V}}^{\left(\delta \right)}$ with price process $V_t=V^{\left(\delta \right)}_t={S_t}^{\delta}$, where $\delta=\frac{-2r}{{\sigma }^2}$.  Hedging strategies involving $\mathcal{M}$  and $\mathbb{V}$ will be presented in the next, Proposition 3.3. 

A natural extension of Merton's model is to assume that the jump dynamics can be potentially dependent on the stock price itself
\begin{equation}
	\label{a3_6}
	{dS}_t= {(}\alpha -\lambda \kappa )S_{t-}dt+\ \sigma S_{t-}{dB}_t+z_t\left({ {y}}_t-1\right)S_{t-}dN_t,\ t\ge 0,
\end{equation}
\begin{equation}
	\label{a3_7}
	dz_t=a^{\left(z\right)}z_tdt+b^{\left(z\right)}z_t{dB}_t\, ,\ a^{\left(z\right)}\in R,\ b^{\left(z\right)}>0
\end{equation}
defined on a stochastic basis $\left( {\Omega },\mathcal{F},{\left\{{\mathcal{F}}_t\right\}}_{t\ge 0} {,}\mathbb{P}\right)$ as in \eqref{a3_1} and \eqref{a3_2}. The risk-free bond dynamics is given by equation \eqref{a2_2}.

Consider a ECC with price process  ${ {Y}}_{ {t}}=Y\left(S_t,z_t,t\right)$, where the function   $Y\left(x,z,t\right)$, $x>0$, $z>0$, $t\ge 0$, is sufficiently smooth. Our goal is to apply a self-financing replicating portfolio which will include (i) the riskless bond , (ii) the stock,  (iii) basic asset  ${\mathbb{V}=\mathbb{V}}^{\left(\delta \right)}$, with price process $V_t=V^{\left(\delta \right)}_t={S_t}^{\delta}$, where $\delta =\frac{-2r}{{\sigma }^2}$, and (iv) basic asset $\mathcal{M}={\mathcal{M}}^{\left(r\right)}$ with price process: $d\mathcal{M}_t=r{\mathcal{M}}_{t-}dt+\ \left({ {y}}_t-1\right){\mathcal{M}}_{t-}dN_t$ $t\ge0$. Then the dynamics of the hedging -portfolio is given by
\begin{equation}
	\label{a3_8}
	dY\left(S_t,z_t,\ t\right)=\ { {c}}^{\left(1\right)}_{ {t}}{dS}_t+{ {c}}^{\left(2\right)}_{ {t}}dS^{\delta }_t+rm_t{d\mathcal{M}}_t+b_t{\beta }_tdt.
\end{equation}

\noindent \textit{Proposition 3.3}. Under the assumptions \eqref{a3_6}, \eqref{a3_7}, and \eqref{a3_7}, $Y\left(x,z,\ t\right)$, $x>0$, $z>0$, $t\ge 0$ satisfies the PDE  (see Appendix A.5 for the proof):

\begin{equation}
	\begin{array}{ll}
		\frac{\partial Y\left( {x,z},t\right)}{\partial t}+r\frac{\partial Y\left( {x,z},t\right)}{\partial x}x+\frac{\partial Y\left( {x,z},t\right)}{\partial z}\left\{\left(\alpha-\lambda \kappa \right)\frac{b^{\left(z\right)}}{\sigma }z-a^{\left(z\right)}\right\}\\
		+\frac{1}{2} \frac{{\partial }^2Y\left(S_{t-},z_t,t\right)}{\partial x^2}{\sigma }^2x^2+\frac{{\partial }^2Y\left(S_{t-},z_t,t\right)}{\partial x\partial z}\sigma b^{\left(z\right)}xz+\frac{1}{2} \frac{{\partial }^2Y\left(S_{t-},z_t,t\right)}{\partial z^2}{b^{\left(z\right)}}^2z^2=0.
	\end{array}
\end{equation} 

\section{Hedging volatility risk in a stochastic volatility option pricing model}
To demonstrate how to hedge the volatility risk in a stochastic volatility option pricing model, we use the same setting as in Section 2. However, in this case we assume that publicly traded stock's price dynamic are determined by a stochastic volatility model with mean-reverting Ornstein--Uhlenbeck process given by \footnote{See section 2.4 in \cite{Fouque:2000}.}

\begin{equation} 
	\label{a4_1} 
	{dS}_t=\mu S_tdt+\ \sigma \left(V_t\right)S_t{dB}_t,\ t\ge 0,\ S_0>0 ,
\end{equation} 
\begin{equation} 
	\label{a4_2} 
	dV_t=\alpha \left(m-V_t\right)dt+\varphi dW_t,\,\,\,\, {dB}_tdW_t=\rho dt, 
\end{equation} 
where   $V_0>0$, $\alpha>0$, $m>0$, $\varphi >0$, $\rho\in \left(-1,1\right)$. We assume that $S_t,\,V_t$ $t\ge 0$ is defined on a stochastic basis $\left( {\Omega },\mathcal{F},{\left\{{\mathcal{F}}_t\right\}}_{t\ge 0},\mathbb{P}\right)$ generated by the correlated Brownian motions $(B_t,W_t)$ $t\ge 0$. The price of the riskless bond is
\begin{equation}
	\label{a4_3}
	\beta _{ {t}}=e^{rt},\ t\ge 0.
\end{equation}
Let ${ {Y}}_{ {t}}=Y\left(S_t,V_t,t\right)$ $t\ge 0$, be the price process of a ECC with maturity $T>0$ and terminal value $Y\left(S_T,V_T,T\right)=\mathcal{G}\left(S_T\right).$  In the standard stochastic volatility model an additional derivative is needed in the replicating portfolio. Then the PDE for the derivative $Y^{\ }_t=Y\left(S_t,V_t,t\right)$ is given by

\begin{equation}
	\label{a4_4}
	\begin{array}{ll}
		\frac{\partial Y\left(x,y,t\right)}{\partial t}+rx\frac{\partial Y\left(x,y,t\right)}{\partial x}+\left(\alpha \left(m-v\right)-\varphi \left(\rho \frac{\mu -r}{\sigma \left(y\right)}+\gamma \left(x,y,t\right)\sqrt{1-{\rho }^2}\right)\frac{\partial Y\left(x,y,t\right)}{\partial y}\right)\\-rY\left(x,v,t\right)
		+\frac{1}{2}\frac{{\partial }^2Y\left(x,y,t\right)}{\partial x^2}{\left(\sigma \left(y\right)x\right)}^2 {+}\frac{ {1}}{ {2}}\frac{{\partial }^2Y\left(x,y,t\right)}{\partial y^2}{\varphi }^2+\frac{{\partial }^2Y\left(x,y,t\right)}{\partial x\partial y}\rho \varphi \sigma \left(y\right)x=0\,,
	\end{array}
\end{equation}
for all $x>0$, $y\in R$, $t\in \left[0,T\right)$, with boundary condition: $Y\left(x,y,T\right)=\mathcal{G}\left(x\right)$ for all $x>0,v\in R$, and $\gamma \left(x,y,t\right)$ being an arbitrary function representing the risk premium factor from the second source of randomness $W_t$ $t\ge 0$.

The objective in this section is to derive a PDE for the price of an ECC in which together with market risk, volatility risk is also expressed in tradeable volatility indexes, and thus the function $\gamma \left(x,y,t\right)$ is removed. 

Next, we follow our basic approach of introducing an additional security as a publicly traded asset to achieve market completeness. Following the approach by \cite{Davis:2004}, we assume that in the stochastic volatility market model, defined by \eqref{a4_1}, \eqref{a4_2}, and \eqref{a4_3}, and consisting of a risky stock and riskless bond, an additional security $\mathcal{V}$ , which we label the ``volatility index", and designated as$\ Mvol$${}^{\ }$  is introduced and publicly traded. The price process $\mathcal{V}$, is the publicly traded price process $V_t$. We assume that $\mathcal{V}$, is a tradable financial instrument such as, for example, the 
futures contract where the underlying is the CBOE Volatility index (VIX)  as a  proxy \citep[see][]{CBOE:2003}.

A trader, having available a traded asset, the bond, the stock and $\mathcal{V},\ $can form a traded portfolio with constant volatility and then apply the ICAPM to optimize his or her wealth process.

First, for a better understanding of the stochastic volatility model we shall prove \eqref{a4_4} using the CCAPM:
\begin{equation}
	\label{a4_5}
	{\mathbb{E}}_t\frac{dS_t}{S_t}=rdt+{\beta }^{\left(S,M\right)}_t\left({\mathbb{E}}_t\ r^{(M)}_t-r\right)dt, 
\end{equation}
\begin{equation}
	\label{a4_6}
	{\mathbb{E}}_t\frac{dY_t}{Y_t}=rdt+{\beta }^{\left(Y,M\right)}_t\left({\mathbb{E}}_t\ r^{(M)}_t-r\right)dt
\end{equation}
where $r^{(M)}_t$ is the market instantaneous return.\footnote{This approach was initially used in \cite{Black:1973} and \citet[Section 6.D]{Duffie:2001}.} We assume that from the available data for stock volatility, the trader estimates the one-factor model for the volatility index security $\mathcal{V}:$
\begin{equation}
	\label{a4_7}
	{\mathbb{E}}_{t}\frac{d{V}_{t}}{{V}_{t}}={\eta }_{t}dt+{\beta }^{\left(V,Mvol\right)}_{t}\left({\mathbb{E}}_{t} {r}^{(Mvol)}_{t}-r\right)dt
\end{equation}

\noindent \textit{Proposition 4.1}. Under the assumptions \eqref{a4_1}, \eqref{a4_2}, \eqref{a4_3}, \eqref{a4_5}, \eqref{a4_6} and \eqref{a4_7}, the ECC-price process $Y_t=Y\left(S_t,V_t,t\right)$ $t\ge0$, has a price dynamics determined by the following PDE for $Y\left(x,y,t\right)$, $x>0$, $z>0$, $t\ge0:$

\begin{equation}
	\label{a4_8}
	\begin{array}{ll}
		\frac{\partial Y\left(x,y,t\right)}{\partial t}+rx\frac{\partial Y\left(x,y,t\right)}{\partial x}+
		\frac{\partial Y\left(x,y,\ t\right)}{\partial y}\left({\eta }_ty+{\beta }^{\left(V,Mvol\right)}_ty{\theta }^{(V)}_t-y{\beta }^{\left(V,M\right)}_t{\theta }^{(M)}_t\right)\\-rY\left(x,y,t\right)
		+\frac{1}{2}\frac{{\partial }^2Y\left(x,y\ t\right)}{\partial x^2}{\left(\sigma \left(y\right)x\right)}^2+\frac{{\partial }^2Y\left(x,y,t\right)}{\partial x\partial y}\sigma \left(y\right)x\varphi \rho +\frac{1}{2}\frac{{\partial }^2Y\left(x,y,\ t\right)}{\partial y^2}{\varphi }^2=0
	\end{array}
\end{equation}
where ${\theta }^{(M)}_t={\mathbb{E}}_t\ r^{\left(Mvol\right)}_t-r$ is the market risk premium, and ${\theta }^{(V)}_t={\mathbb{E}}_t\ r^{\left(Mvol\right)}_t-r$ is the volatility risk premium  (see Appendix A.6 for the proof).

First, it should be noted, that if the stock's volatility is a security $\mathcal{V}$  with publicly traded price process $V_t,$ then a trader should apply a self-financing strategy 
\begin{equation}
	\label{a4_9}
	Y_{t}=a_tS_t+b_{{t}}{{\beta }}_{t}+{c}_{t}V_t
\end{equation}
with 
\begin{equation}
	\label{a4_10} 
	\left\lbrace \begin{array}{lll}
		a_t=\frac{\partial Y\left(S_t,V_t,t\right)}{\partial x},\\
		{b}_{t}=\frac{1}{{{\beta }}_{t}}\left\{Y\left(S_t,V_t,t\right)-\frac{\partial Y\left(S_t,V_t,t\right)}{\partial x}S_t-\frac{\partial Y\left(S_t,V_t,t\right)}{\partial y}V_t\right\},\\
		{c}_{t}=\frac{\partial Y\left(S_t,V_t,t\right)}{\partial y}.   
	\end{array}  
	\right.       
\end{equation} 
\textit{Proposition 4.2} \cite{Davis:2004}. Under the assumptions \eqref{a4_1}, \eqref{a4_2}, \eqref{a4_3}, \eqref{a4_9} and \eqref{a4_10}, the ECC-price process $Y_t=Y\left(S_t,V_t,t\right)$ $t\ge 0$ has price dynamics determined by the following PDE for $Y\left(x,y,t\right)$, $x>0$, $z>0$, $t\ge 0$:

\begin{equation}
	\begin{array}{ll}
		\frac{\partial Y\left({x,y},t\right)}{\partial t}+r\frac{\partial Y\left({x,y},t\right)}{\partial x}{x+}\frac{\partial Y\left({x,y}, t\right)}{\partial y}y-rY\left(x,y,t\right)+\frac{1}{2}\frac{{\partial }^2Y\left({x,y},t\right)}{\partial x^2}{\left(\sigma \left({y}\right){x}\right)}^2\\+ 
		\frac{{\partial }^2Y\left({x,y},t\right)}{\partial x\partial y}\sigma \left({y}\right){x}\varphi \rho +\frac{1}{2}\frac{{\partial }^2Y\left({x,y}, t\right)}{\partial y^2}{\varphi }^2=0.
	\end{array}
\end{equation}
The proof of the proposition follows the same arguments as Proposition 3.3 \footnote{ See \cite{Davis:2004}.} and thus is omitted.

Within the stochastic volatility model suggested by \cite{Heston:1993}, if  $\mathcal{V}$ is a publicly traded asset, that will allow the trader to form a synthetic self-financing portfolio ${\mathcal{P}}^{\left(S,\mathcal{V}\right)}\ $of the stock and  $\mathcal{V}$ so that  ${\mathcal{P}}^{\left(S,\mathcal{V}\right)}$ has constant volatility. As a second step the investor can use Merton's ICAPM to form an dynamically optimal wealth portfolio.

Proposition 4.1 leads to the following generalization of the stochastic volatility model. Let us remark here, that trading volatility options is gaining popularity, and involves models for volatility options in which the ``volatility of the volatility" (designated as vol-of-vol, or shortly, $\mathcal{V}o\mathcal{V}$) must be modeled. We now introduce a stock-price model (designated as $\mathcal{V}o\mathcal{V}$-model) in which the volatility and the $\mathcal{V}o\mathcal{V}$ are It\^{o} processes:
\begin{equation}
	\label{a4_11}
	\left\lbrace \begin{array}{llll}
		{dS}_t={\mu }_tS_tdt+\ \sigma \left(V_t\right)S_t{dB}_t\,, \\
		{dV}_t={\alpha }_tdt+\varphi \left({{v}}_t\right)dB^{\left(V\right)}_t,\\
		d{{v}}_t=b_tdt+{\psi }_tdB^{\left({v}\right)}_t,\\
		{dB}_tdB^{\left(V\right)}_t={\rho}^{\left(V\right)}dt,\,\,\,{dB}_tdB^{\left({v}\right)}_t={\rho}^{\left({v}\right)}dt,\,\,\, dB^{\left(V\right)}_tB^{\left({v}\right)}_t=\varrho dt,\\ S_0>0,\,\,V_0>0,\,\,{{v}}_0>0,\,\,t\ge 0,\,\, {\rho }^{\left(V\right)},{\rho }^{\left({v}\right)},\varrho \in \left(-1,1\right).
	\end{array}
	\right.
\end{equation}

The $\mathcal{V}o\mathcal{V}$-model is sufficiently flexible to capture volatility clustering of the returns and a second-order volatility clustering (``volatility clustering of the volatility'', or what can be called the \textit{roughness} of the price process). 

Let ${{Y}}_{{t}}=Y\left(S_t,V_t,{{v}}_t,t\right)$, $t\ge 0$, be the price process of a ECC with maturity $T>0$ and terminal value $Y\left(S_T,V_T,T\right)=\mathcal{G}\left(S_T\right)$.  We first assume that security  $\mathcal{V}$ with price process  $V_t$ $t\ge 0$, is publicly traded.  Second, following our approach toward market completeness, we assume that the market has introduced  $\mathcal{V}o\mathcal{V}$. The security $\mathcal{V}o\mathcal{V}$ is publicly traded with price process  ${{v}}_t$, $t\ge0$, having price dynamics given in \eqref{a3_1}.\footnote{ We view VIX (an index created by CBOE, representing 30-day implied volatility calculated by S\&P 500 options) as a proxy for publicly traded security $\mathcal{V}$ (see\ http://www.cboe.com/vix). As for $\mathcal{V}o\mathcal{V}$ we do not have an existing market proxy. However, the financial industry is already trying to construct a synthetic security which mimics $\mathcal{V}o\mathcal{V}$-dynamics, and, it is our belief, that it should not be long before such a ``vol-of-vol-index" is introduced as a publicly traded asset.} Then applying It\^{o}'s formula and considering the self-financing strategy ${{Y}}_{{t}}=Y\left(S_t,V_t,{{v}}_t,t\right)={{\Delta }}^{\left(S\right)}_t{dS}_t+{{\Delta }}^{\left(V\right)}_t{dV}_t+{{\Delta }}^{\left({v}\right)}_t{d{v}}_t+{{b}}_{{t}}{{\beta }}_{ {t}}$, where ${ {\beta}}_{ {t}}=e^{rt}$, $t\ge 0$, is the riskless bond, results in

\begin{equation}
	{ {\Delta }}^{\left(S\right)}_t=\frac{\partial Y\left(S_t,V_t, { {v}}_t,t\right)}{\partial x},\ { {\Delta }}^{\left(V\right)}_t=\frac{\partial Y\left(S_t,V_t,{ {v}}_t,t\right)}{\partial y},\,{ {\Delta }}^{\left( {v}\right)}_t=\frac{\partial Y\left(S_t,V_t,{ {v}}_t,t\right)}{\partial z}.
\end{equation}

Following standard no-arbitrage arguments leads to the PDE for the ECC-price process:
\begin{equation}
	\begin{array}{lll}
		\frac{\partial Y\left( {x,y,z},t\right)}{\partial t}+\frac{\partial Y\left(x,y,z,t\right)}{\partial x}rx+\frac{\partial Y\left(x, {y},z,t\right)}{\partial y}ry+\frac{\partial Y\left(x,y,z,t\right)}{\partial z}rz-rY\left(x,y,z,t\right) 
		\\+\frac{1}{2}\frac{{\partial }^2Y\left(x,y,z,t\right)}{\partial x^2}{\left(\sigma \left( {y}\right)x\right)}^2+\frac{1}{2}\frac{{\partial }^2Y\left(x,y,z,t\right)}{\partial x^2}{\left(\sigma \left( {y}\right)x\right)}^2+\frac{1}{2}\frac{{\partial }^2Y\left(x,y,z,t\right)}{\partial y^2}{\left(\varphi \left(z\right) {y}\right)}^2 
		\\+\frac{1}{2}\frac{{\partial }^2Y\left(x,y,z,t\right)}{\partial z^2}{\left({\psi }_tz\right)}^2+\frac{1}{2}\frac{{\partial }^2Y\left(x,y,z,t\right)}{\partial z^2}{\left({\psi }_tz\right)}^2+\frac{{\partial }^2Y\left(x,y,z,t\right)}{\partial x\partial y}{\rho }^{\left(V\right)}\sigma \left(y\right)x\varphi \left({ {v}}_t\right)V_t 
		\\+\frac{{\partial }^2Y\left(S_t,V_t,{ {v}}_t,t\right)}{\partial x\partial z}{\rho }^{\left( {v}\right)}\sigma \left(V_t\right)S_t{\psi }_t{ {v}}_t+\frac{{\partial }^2Y\left(S_t,V_t,{ {v}}_t,t\right)}{\partial y\partial z}\varrho \varphi \left({ {v}}_t\right)V_t{\psi }_t{ {v}}_t=0.
	\end{array}
\end{equation}

Next we can extend the example of the use of CCAPM in the previous section by applying CCAPM not only to the stock, derivative, and the volatility security $\mathcal{V},$ but to the new security $\mathcal{V}o\mathcal{V}$ as well:
\begin{equation}
	\begin{array}{llll}
		{\mathbb{E}}_t\frac{dS_t}{S_t}=rdt+{\beta }^{\left(S,M\right)}_t\left({\mathbb{E}}_t\ r^{\left(M\right)}_t-r\right)dt,\\ 
		{\mathbb{E}}_t\frac{dY_t}{Y_t}=rdt+{\beta }^{\left(Y,M\right)}_t\left({\mathbb{E}}_t\ r^{\left(M\right)}_t-r\right)dt,\\ 
		{\mathbb{E}}_t\frac{dV_t}{V_t}={\eta }^{\left(V\right)}_tdt+{\beta }^{\left(V,Mvol\right)}_t\left({\mathbb{E}}_t\ r^{(Mvol)}_t-r\right)dt,\\ 
		{\mathbb{E}}_t\frac{d{ {v}}_t}{{ {v}}_t}={\eta }^{\left( {v}\right)}_tdt+{\beta }^{\left( {v},Mvol\right)}_t\left({\mathbb{E}}_t\ r^{(Vvol)}_t-r\right)dt. 
	\end{array}
\end{equation}

Applying the same arguments as in the previous section results in the following PDE:
\begin{equation}
	\label{vol_vol}
	\begin{array}{llll}
		\frac{\partial Y\left(x,y,z,t\right)}{\partial t}+\frac{\partial Y\left(x,y,z,t\right)}{\partial x}rx+\frac{\partial Y\left(x,y,z,t\right)}{\partial y}\left({\eta }^{\left(V\right)}_t {y}+{\beta }^{\left(V,Mvol\right)}_t {y}{\theta }^{\left(V\right)}_t-y{\beta }^{\left(V,M\right)}_t{\theta }^{(M)}_t\right)\\ 
		+\frac{\partial Y\left(S_t,V_t\ ,\ { {v}}_t,t\right)}{\partial z}\left({\eta }^{\left( {v}\right)}_t{ {v}}_t+{\beta }^{\left( {v},Mvol\right)}_t{ {v}}_t{\theta }^{\left( {v}\right)}_t-z{\beta }^{(V, {v})}_t{\theta }^{(M)}_t\right)-rY\left(x,y,z\right) 
		\\+\frac{1}{2}\frac{{\partial }^2Y\left(x,y,z,t\right)}{\partial x^2}{\left(\sigma \left( {y}\right) {x}\right)}^2
		+\frac{1}{2}\frac{{\partial }^2Y\left(x,y,z,t\right)}{\partial y^2}{\left(\varphi \left(z\right)y\right)}^2+\frac{1}{2}\frac{{\partial }^2Y\left(x,y,z,t\right)}{\partial z^2}{\left({\psi }_tz\right)}^2 
		\\+\frac{{\partial }^2Y\left(x,y,z,t\right)}{\partial x\partial y}{\rho }^{\left(V\right)}\sigma \left(y\right)x\varphi \left(z\right)y+\frac{{\partial }^2Y\left(x,y,z,t\right)}{\partial x\partial z}{\rho }^{\left( {v}\right)}\sigma \left(y\right)x{\psi }_tz 
		\\+\frac{{\partial }^2Y\left(x,y,z,t\right)}{\partial y\partial z}\varrho \varphi \left(z\right)y{\psi }_tz=0,
	\end{array}
\end{equation}
where ${\theta }^{(M)}_t={\mathbb{E}}_t\ r^{\left(Mvol\right)}_t-r$ is the market risk premium,  ${\theta }^{(V)}_t={\mathbb{E}}_t\ r^{\left(Mvol\right)}_t-r$ is the volatility risk premium, and ${\theta }^{( {v})}_t={\mathbb{E}}_t\ r^{\left(Vvol\right)}_t-r$  is the vol-of-vol premium.

It is natural to seek an extension of stochastic volatility models \eqref{vol_vol} allowing for jumps in the stock price. We suggest the following one:
\begin{equation}
	\label{volatility}
	\noindent  \left\lbrace \begin{array}{llll}
		{dS}_t=\left(\alpha -\lambda \kappa \right)S_{t-}dt+\ \ \sigma \left(V_t\right)S_{t-}{dB}_t+z_t\left({ {y}}_t-1\right)S_{t-}dN_t,\\ 
		{dz}_t=a^{\left(z\right)}z_tdt+b^{\left(z\right)}z_t{dB}_t,\\ 
		{dV}_t=\alpha \left(m-V_t\right)dt+\varphi dW_t\,, \,\,\,\,{dB}_tdW_t=\rho dt,\\
		t\ge 0,\, a^{\left(z\right)}\in R,\ b^{\left(z\right)}>0, V_0>0,\ \alpha >0,\, m>0,\ \varphi >0,\ \rho \in \left(-1,1\right). 
	\end{array}
	\right.
\end{equation}

The triplet $\left(S_t,z_t,\ V_t\right),t\ge 0,\ $is defined on a stochastic basis $\left( {\Omega },\mathcal{F},{\left\{{\mathcal{F}}_t\right\}}_{t\ge 0} {,}\mathbb{P}\right)$ generated by the correlated Brownian motions $B_t$ $W_t$ and Poisson process $N_t$ and an independent of 

\noindent $\left(B_t,W_t,N_t \right)$, jump-amounts (jump-sizes) 
\begin{equation}
	J^{\left(l\right)}\triangleq J\triangleq \left\{ \begin{array}{c}
		ln\psi \ \ \ \ w.p.\ \ p\in (0,1) \\ 
		0\ \ \ \ \ \ \ w.p.\ \ \ \ \ 1-p \end{array}
	\right.
\end{equation}
The riskless bond dynamics are given by equation \eqref{a2_2}.

Then following the definitions of the new publicly traded assets introduced in this and the previous section, we define a hedging portfolio with the following dynamics: 
\[dY\left(S_t,z_t,t\right)={ {c}}^{\left(1\right)}_{ {t}}{dS}_t+{ {c}}^{\left(2\right)}_{ {t}}dS^{\delta }_t+{ {c}}_{ {t}}V_t+rm_t{d\mathcal{M}}_t+b_t{\beta }_tdt,\] 
where $ {d\mathcal{M}}_t={ {d}\mathcal{M}}_{ {t}} {=r}{\mathcal{M}}_{ {t-}} {dt+}\left({ {y}}_{ {t}} {-} {1}\right){\mathcal{M}}_{ {t-}} {d}{ {N}}_{ {t}}$ $t\ge0$.

The corresponding PDE for $Y\left( {x},z,t\right)$, $x>0$, $z>0$, $t\ge0$, can be readily derived following the same arguments as the proof of Proposition 3.3.

\section{Empirical analysis}
In this section we apply the new trading instrument we proposed in this paper to explain and evaluate market risk before and during an actual  distressed market period. A standard risk measure  must be employed to measure financial industries and market risk. The two measures we selected are Value at Risk (VaR) and Conditional VaR (CVaR) because they are the  two most popular risk measures used in the finance industry. CVaR, also called expected tail risk, is the average of VaRs less than the VaR for a given tail probability. CVaR satisfies all attributes of a coherent risk measure and is consistent with performance relations of risk-averse investors \citep[see][]{Pflug:2000}. 

TLR has been adapted as the risk measure to determine financial market behavior before and during the high volatility period we study later in this paper. For a given tail probability $\alpha$, we define TLR$\left(\alpha \right)$ as follows
\begin{equation}
	TLR\left(\alpha \right) =\frac{CVaR(\alpha)-VaR(\alpha)}{VaR(\alpha)}\,,\alpha \in \left(0,1 \right). 
\end{equation}
The idea of tail-loss ratio is that the tail behavior of the return-distribution measured by the tail-loss can be used as predictor of future market crashes. \cite{kim:2011} discussed a particular dynamic model, ARMA(1,1)-GARCH(1,1) with tempered stable innovations (ARMA-GARCH-TS), for the stock price. They defined tail loss as the differences between $99\%$ CVaR for the ARMA-GARCH-normal and $99\%$ CVaR for the ARMA-GARCH-TS models and found that tail-losses can be used as an early warning system for a forthcoming sharp market downturn. We decided to use a model-free tail loss by using historical returns only for the stock price and the new security. We want to see whether the tail-losses provide some early insight about a future crash. 

Of interest is the perpetual derivative   ${\mathbb{V}=\mathbb{V}}^{\left(\delta \right)}$, where $\delta =\frac{-2r}{{\sigma}^2}$, with price process   $V_t=V^{\left(\delta\right)}_t={S_t}^{\delta}$. To form the price process of $V_t$ $t\geq0$ , we use market indices by the triplet $\left(S_t,\, \sigma_t,\, r_t \right)$ $t\geq0$ where: (i) $S_t$ $t\geq0$ as the price process of the SPDR S\&P 500 index; (ii) $\sigma_t$ $t\geq0$ as the cumulative VIX (i.e., $\sigma_t$ represent the cumulative of VIX in $[0,t]$), and (iii) $r_t$ $t\geq0$ is the  10-year Treasury yield published by the US Treasury on its website \url{https://home.treasury.gov/}. The database to form the time series of $V_t$ covers the period from January 1993 to June 2019. There were including 6661 observations collected from \textit{Yahoo Finance}. We selected a broad-based market index, the S\&P 500, as measured by the SPDR S\&P 500 $\left(S_t, \,t\geq0 \right)$ which is an exchange-traded index as a benchmark to compare it with our new financial instrument during the market financial crisis. We evaluate the market risk prior to the distressed period investigated and then assess whether the tail-losses provide some early insight about the actual market crash. 

\begin{figure}[t!]
	\centering
	\includegraphics[width=13cm, height=8cm]{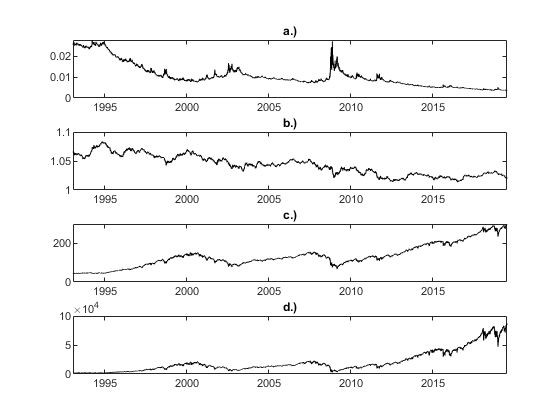}
	\caption{Daily time series process for the period from 
		January 1993 to June 2019  for (a) $V^{\left(-1\right)}_t$ the basic asset of order -1; 
		(b) $V^{\left(0\right)}_t$ the riskless bond;
		(c) $V^{\left(1\right)}_t$ the risky asset (S\&P 500 index closing price, and;  
		(d) $V^{\left(2\right)}_t$ the basic asset of order 1. }
	\label{Fig_4}
\end{figure}

In Figure \ref{Fig_4} the price process of the perpetual derivative $V^{\left(n\right)}_t$ (defined in Proposition 1) for $n=\left\lbrace -1,0,1,2\right\rbrace $ is plotted. We designated $V^{(0)}$ as a basic asset of order 0 (riskless asset) and $V^{(1)}$ as a basic asset of order 1 (risky asset). $V^{(-1)}$ and $V^{(1)}$ would be viewed as a basic asset of order -1 and 1, respectively.

\begin{figure}[t!]
	\centering
	\includegraphics[width=13cm, height=8cm]{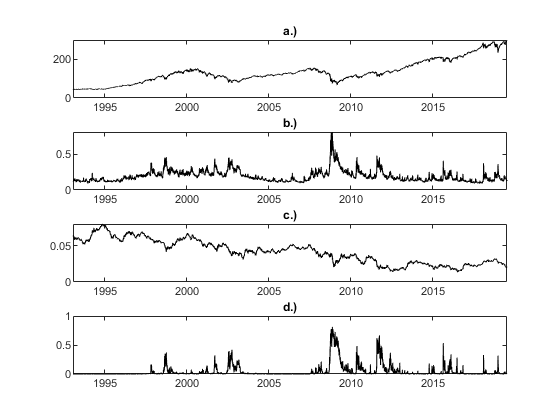}
	\caption{Daily time series process for the period from 
		January 1993 to June 2019  for (a) the SPDR S\&P 500 index closing price $\left( S_t\right) $; 
		(b) the VIX index closing price $\left( \sigma_t\right) $;
		(c) 10-year Treasury yield $\left( r_t\right)$, and; 
		(d) the perpetual derivative $\left( V_t\right)$. }
	\label{Fig_1}
\end{figure}

Figure \ref{Fig_1} shows the time series price process of $V_t$ corresponding to daily price process for the $S_t$ index and VIX index based on the closing price and the 10-year Treasury yield. The stable and small volatility values of $V_t$   observed for a few months preceding the 2008 market crash indicated an early warning for a pending market crash. The volatility of $V_t$ increases while the VIX index falls. The apparent volatility of $V_t$ is higher in the prior four years than in the years preceding the financial crises. 

Next we calculate CVaR and VaR at $\alpha=99\%$ confidence level and thereafter obtain the TLR at $99\%$ confidence level for log-return of $V_t$ and $S_t$ indices. We then analyze the market before and during the 2008 market crash.

\begin{figure}[t!]
	\centering
	\includegraphics[width=13cm, height=10cm]{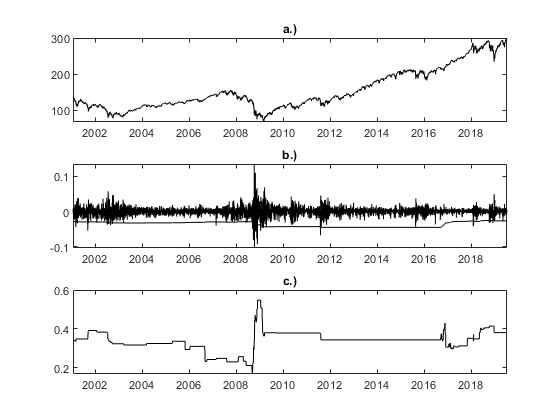}
	\caption{Daily time series process for the period from 
		January 2001 to June 2019  for (a) the price process of SPDR S\&P 500 index ; 
		(b) the log-return of SPDR S\&P 500 index with emperical VaR(0.99), and; 
		(c) the TLR index for SPDR S\&P 500 index.
	}
	\label{Fig_2}
\end{figure}

\begin{figure}[b!]
	\centering
	\includegraphics[width=13cm, height=10cm]{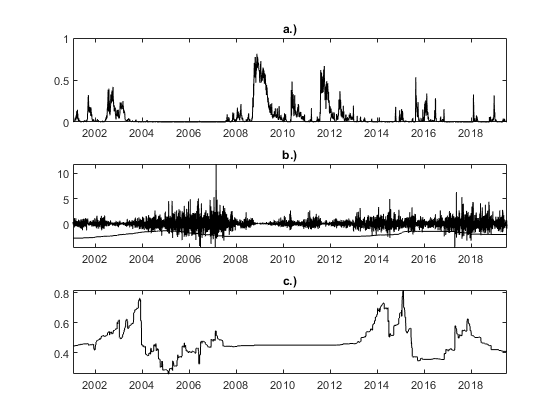}
	\caption{Daily time series process for the period from 
		January 2001 to June 2019  for (a) the price process of the perpetual derivative index; 
		(b) the log-return of the perpetual derivative index with emperical VaR(0.99), and;
		(c) the TLR index for the perpetual derivative index.
	}
	\label{Fig_3}
\end{figure}

Empirical daily VaR and CVaR are estimated by using rolling windows time series analysis for a time period of eight years (i.e. 2016 trading days). 
Then, the TLR index for the daily log-return of $S_t$ and $V_t$ are calculated. 
The TLR indices associated with $S_t$ and $V_t$ are shown in Figure \ref{Fig_2}  and Figure \ref{Fig_3} , respectively.
Comparing Figure \ref{Fig_2}  to Figure \ref{Fig_3}, one can graphically check the difference between the two TLR indices. In particular, the TLR index of $S_t$ seems to be more volatile during the 2008 market crash. The index has low volatility from July 2007 to May 2008 and start slowly declining thereafter. Finally, it sharply increase in September 2008. Because no specific pattern is observed, we cannot observe any early warning indicator of a pending market crash based on the TLR index of $S_t$. 

The TRS index of $V_t$ begins to stabilize with low volatility a few months preceding the market crash of 2008. It remains stable during the market crash, while it can be seen that there is high volatility in the the TRS index of $S_t$ during the 2008 market crash. We consider this low volatility as an early warning indicator of a pending market crash. Due to the presence of long memory in stock market the volatility of the TLR  remains stable after the market crash.  Hence, we consider the low volatility of the TLR index for $V_t$ as a good measure to indicate a forthcoming sharp market downturn.

\section{Conclusions}  

In this paper, we developed a novel approach to hedging derivatives by proposing the creation of a set of new specially designed (``ideal'') publicly available perpetual derivatives. The set of new proposed assets - perpetual options - allow the option writer to form a hedging portfolio which (1) reduces the trading costs in markets with frictions; (2) removes the jump risk when the underlying price process is a jump-diffusion process, and; (3) removes the volatility risk when the underlying price process exhibits stochastic volatility. The corresponding PIDE and PDE for the derivative price processes are derived. The tail-loss ratio is used as the risk measure to determine financial market behavior prior to the market high volatility period we study in this paper. We demonstrate that the new financial instruments, together with the tail-loss ratio have potential  power in predicting and evaluating market risk before a distressed market period. Our findings suggest that the tail-loss ratio can be potentially used as a metric for an early warning system.

\newpage
\section*{Appendix}
\indent
\begin{appendix}
	\appendix{\textbf{A.1. Proof of Proposition 2.1}}	
	\label{appendix:i}
	
	\noindent   Let $V_t=g\left(S_t,{\beta }_t\right)$ where $g\left(x,y\right)$, $x\ge 0$, $y\ge 0$ is a sufficiently smooth function. Then  $\frac{V_t}{{\beta }_t}$ is a $\mathbb{Q} {-}$martingale if and only if  $\frac{\partial }{\partial x}g\left(x,y\right)rx+\frac{\partial }{\partial y}g\left(x,y\right)ry-rg\left(x,y\right)+\frac{1}{2}\frac{{\partial }^2}{\partial x^2}g\left(x,y\right){\sigma }^2x^2=0$, which is satisfied for $g\left(x,y\right)=x^{\zeta }y^{\gamma}$.\\
	
\end{appendix}

\begin{appendix}
	\appendix{\textbf{A.2. Proof of Proposition 2.2}}	
	\label{appendix:ii}
	
	\noindent For simplicity of the exposition we will consider the bivariate case only. First let us recall the Black-Scholes Equation in the bivariate case: The stocks dynamics is given by $dX_t={\mu }^{(X)}_tX_tdt+{\sigma }^{\left(X,B\right)}_tX_tdB_t+{\sigma }^{\left(X,W\right)}_tX_tdW_t$, $dY_t={\mu }^{(Y)}_tY_tdt+{\sigma }^{\left(Y,B\right)}_tY_tdB_t+{\sigma }^{\left(Y,W\right)}_tY_tdW_t$, and given an ECC with price process $C_t=C\left(X_t,Y_t,t\right).$ We replicate  $C_t=C(X_t,Y_t,t)$ by a self-financing strategy  $C\left(X_t,Y_t,t\right)=a^{(X)}_tX_t+a^{(Y)}_tY_t+b_t{\beta }_t$. Then
	\[dC\left(X_t,Y_t,t\right)=a^{(X)}_t{dX}_t+a^{(Y)}_t{dY}_t+b_td{\beta }_t\] 
	\[=dt+\left(a^{\left(X\right)}_t{\sigma }^{\left(X,B\right)}_tX_t+a^{\left(Y\right)}_t{\sigma }^{\left(Y,B\right)}_tY_t\right)dB_t+\left(a^{\left(X\right)}_t{\sigma }^{\left(X,W\right)}_tX_t+a^{\left(Y\right)}_t{\sigma }^{\left(Y,W\right)}_tY_t\right)dW_t.\] 
	Equating the terms for $dC\left(X_t,Y_t,t\right)$ leads to $a^{\left(X\right)}_t=\frac{\partial C\left(X_t,Y_t,t\right)}{\partial x}$, $a^{\left(Y\right)}_t=\frac{\partial C\left(X_t,Y_t,t\right)}{\partial y}$, and the no-arbitrage assumption holds if and only if $\delta =-\frac{2r}{{\left({\sigma }^{\left(X,B\right)}_t\right)}^2+{\left({\sigma }^{\left(X,W\right)}_t\right)}^2}$, and $\gamma=-\frac{2r}{{\left({\sigma }^{\left(Y,B\right)}_t\right)}^2+{\left({\sigma }^{\left(,W\right)}_t\right)}^2}$, which completes the proof of the proposition.\\
	
\end{appendix}

\begin{appendix}
	\appendix{\textbf{A.3. Proof of Proposition 3.1}}	
	\label{appendix:iii}
	
	\noindent Applying CCAPM and making use of Merton's assumption\\ ${\mathbb{E}}_tr^{(S)}_t=r-{\beta }^{(S,M)}_t\left({\mathbb{E}}_tr^{(M)}_t-r\right)$, ${\mathbb{E}}_t r^{(Y)}_t=r-{\beta }^{(Y,M)}_t\left({\mathbb{E}}_tr^{(M)}_t-r\right)$,\\  ${\beta }^{(Y,M)}_t=\frac{{{Cov}_t(r}^{\left(Y\right)}_t,r^{\left(M\right)}_t)}{{var}_t\left(r^{\left(M\right)}_t\right)}=\frac{\partial Y\left(S_t,t\right)}{\partial x}\frac{S_t}{Y_t}{\beta}^{(S,M)}_t$,\\ we have two expressions for ${{\mathbb{E}}_t {dY}}_{ {t}}={ {Y}}_{ {t}}{\mathbb{E}}_tr^{(Y)}_tdt$:\\
	${{\mathbb{E}}_t {dY}}_{ {t}}$=$\left(\frac{\partial Y\left(S_{t-},t\right)}{\partial t}+\frac{1}{2}{\sigma }^2S^2_{t-}\frac{{\partial }^2Y\left(S_t,t\right)}{{\partial x}^2}\right)dt$+$ \frac{\partial Y\left(S_t,t\right)}{\partial x}\ S_t\left(r-{\beta }^{\left(S,M\right)}\left({\mathbb{E}}_tr^{\left(M\right)}_t-r\right)\right)dt$
	\\$+\left\{\mathbb{E}\left[Y\left({{ {y}}_tS}_t,t\right)-Y\left(S_{t-},t\right)-\frac{\partial Y\left(S_t,t\right)}{\partial x}S_{t-}({ {y}}_t-1)\right]\right\}\lambda dt$\\
	and  ${{\mathbb{E}}_t {dY}}_{ {t}}$$=rY_tdt-\frac{\partial Y\left(S_t,t\right)}{\partial x}S_t{\beta }^{(S,M)}_t\left({\mathbb{E}}_tr^{\left(M\right)}_t-r\right)dt$ . 
	Equating  ${{\mathbb{E}}_t {dY}}_{ {t}}$ terms and setting $S_t=x\ $completes the proof of the proposition.\\	
\end{appendix}

\begin{appendix}
	\appendix{\textbf{A.4. Proof of Proposition 3.2}}	
	\label{appendix:iv}
	
	\noindent Consider the self-financing trading strategy  $Y\left(S_t,M_t,t\right)=a_tS_t{+c}_tM_t+b_t{\beta }_t$.  Equating the terms for $dY\left(S_t,M_t,t\right)$ leads to: ${\ a}_t=\frac{\partial Y\left(S_{t-},M_t,t\right)}{\partial x}$,  and
	$c_t=\frac{\partial Y\left(S_{t-},M_t,t\right)}{\partial z}-\frac{S_{t-}}{M_{t-}}-\frac{\partial Y\left(S_{t-},M_t,t\right)}{\partial x}\frac{Y\left(S_t,M_t,\ t\right)-Y\left(S_{t-},M_{t-},t\right)}{\left(\psi -1\right)M_{t-}}.$ 
	Comparing the term $\left(\dots \right)dt\ $ completes the proof of the proposition.\\
	
\end{appendix}	

\begin{appendix}
	\appendix{\textbf{A.5. Proof of Proposition 3.3}}	
	\label{appendix:v}
	
	\noindent We seek a self-financing portfolio of the form $Y\left(S_t,z_t,\ t\right)=\ { {c}}^{\left(1\right)}_{ {t}}S_t+{ {c}}^{\left(2\right)}_{ {t}}S^{\delta }_t+m_t{\mathcal{M}}_t+b_t{\beta }_t$ with  ${ {dY}}_{ {t}}=dY\left(S_t,z_t,\ t\right)=\ { {c}}^{\left(1\right)}_{ {t}}{dS}_t+{ {c}}^{\left(2\right)}_{ {t}}{dS}^{\delta }_t+m_t{d\mathcal{M}}_t+b_t{d\beta }_t$. To capture the ${dz}_t$-risk, we use the replicating portfolio the representation:
	${dS}_t=\left\{\left(\alpha -\lambda \kappa \right)S_{t-}-\sigma S_{t-}\frac{a^{\left(z\right)}}{b^{\left(z\right)}}\frac{1}{z_t}\right\}dt+\sigma S_{t-}\frac{1}{b^{\left(z\right)}}{\frac{1}{z_t}dz}_t+z_t\left({ {y}}_t-1\right)S_{t-}dN_t$ 
	in the term ${ {c}}^{\left(1\right)}_{ {t}}{dS}_t$. Thus \\
	$m_t=\frac{\left(\psi S_tz_t,\ t\right)-Y\left(S_{t-}z_t,t\right)}{\left(\psi -1\right){\mathcal{M}}_{t-}}-\frac{\partial Y\left(S_{t-},z_t,\ t\right)}{\partial z}\frac{b^{\left(z\right)}z_t}{\sigma }\frac{1}{{\mathcal{M}}_{t-}}-\frac{1}{\delta }\frac{\partial Y\left(S_{t-},z_t,\ t\right)}{\partial x}S_{t-}{z_t}^{\delta }\frac{\left[{ {\psi }}^{\delta }-1\right]}{\left(\psi -1\right){\mathcal{M}}_{t-}}.$ 
	With similar expressions for ${ {c}}^{\left(1\right)}_{ {t}}$ and ${ {c}}^{\left(2\right)\ }_{ {t}}$. Then equalizing he terms with $\left(\dots \right)dt$ results in the required PDE, completing the proof of the proposition.\\
	
\end{appendix}

\begin{appendix}
	\appendix{\textbf{A.6. Proof of Proposition 4.1}}	
	\label{appendix:vi}
	
	\noindent Combining the expressions for\\
	${\mathbb{E}}_td{ {Y}}_{ {t}}$=$\left\{ \begin{array}{c}
	\frac{\partial Y\left(S_t,V_t\ t\right)}{\partial t}  
	+\frac{1}{2}\frac{{\partial }^2Y\left(S_t,V_t\ t\right)}{\partial x^2}{\left(\sigma \left(V_t\right)S_t\right)}^2\\+\frac{{\partial }^2Y\left(S_t,V_t\ t\right)}{\partial x\partial y}\sigma \left(V_t\right)S_t\varphi \rho  
	+\frac{1}{2}\frac{{\partial }^2Y\left(S_t,V_t\ t\right)}{\partial y^2}{\varphi }^2 \end{array}
	\right\}dt$+\\$\frac{\partial Y\left(S_t,V_t\ t\right)}{\partial x}\left(S_trdt+{\beta }^{\left(S,M\right)}_tS_t\left({\mathbb{E}}_t\ r^{\left(M\right)}_t-r\right)\right)dt$+\\$\frac{\partial Y\left(S_t,V_t\ t\right)}{\partial y}{\mathbb{E}}_t\left({\eta }_tV_tdt+{\beta }^{\left(V,Mvol\right)}_tV_t\left({\mathbb{E}}_t\ r^{(Mvol)}_t-r\right)\right),$\\
	and \\${\mathbb{E}}_td{ {Y}}_{ {t}}=rY_tdt+\left\{\frac{\partial Y\left(S_t,V_t,t\right)}{\partial x}S_t{\beta }^{(S,M)}_t+\frac{\partial Y\left(S_t,V_t,t\right)}{\partial y}V_t{\beta }^{(V,M)}_t\right\}\left({\mathbb{E}}_t\ r^{(M)}_t-r\right)dt$ \\with \\ $Y_t{\beta }^{(Y,M)}_t={Cov(r}^{\left(Y\right)}_t,\ r^{\left(M\right)}_t)=\frac{\partial Y\left(S_t,V_t,t\right)}{\partial x}S_t{\beta }^{(S,M)}_t+\frac{\partial Y\left(S_t,V_t,t\right)}{\partial y}V_t{\beta }^{(V,M)}_t$\\ which proves the proposition. \\
	
\end{appendix}

\normalem

\end{spacing}
\end{document}